\documentclass[twoside,web]{ieeecolor}
\usepackage{generic}
\usepackage{cite}
\usepackage{amsmath,amssymb,amsfonts}
\usepackage{algorithmic}
\usepackage{graphicx}
\usepackage{textcomp}
\def\BibTeX{{\rm B\kern-.05em{\sc i\kern-.025em b}\kern-.08em
    T\kern-.1667em\lower.7ex\hbox{E}\kern-.125emX}}
\begin{document}

\title{Content Addressable Memory Design with Reference Resistor for Improved Search Resolution}

\author{Siri~Narla,\thanks{This work was financially supported by CoCoSys, part of the JUMP 2.0 program, sponsored by DARPA, University of Michigan College of Engineering,  and Intel Corporation, and was technically supported by the Michigan Center for Materials Characterization, University of Michigan Lurie Nanofabrication Facility and the Minnesota Nano Center, supported by the NSF NNCI (Award ECCS-2025124).}~Ruixue~Li,~Steven~J.~Koester,~\IEEEmembership{~Fellow,~IEEE}, ~Rebecca~A.~Dawley,~Ageeth~A.~Bol,~Piyush~Kumar,~\IEEEmembership{Student~Member,~IEEE},~and~Azad~Naeemi,~\IEEEmembership{Senior~Member,~IEEE}}

\maketitle

\begin{abstract}
Despite the parallel in-memory search capabilities of content addressable memories (CAMs), their use in applications is constrained by their limited resolution that worsens as they are scaled to larger arrays or advanced nodes. In this work we present experimental results for a novel back-end-of-line compatible reference resistive device that can significantly improve the search resolution of CAMs implemented with CMOS and beyond-CMOS technologies to $\leq$ 5-bits.

\end{abstract}
\section{Introduction}
Performing accurate and fast similarity search has become a crucial step for many artificial intelligence (AI) applications such as recommendation systems \cite{sasrec}, retrieval tasks \cite{Noh_2017_ICCV} and hyperdimensional compute tasks \cite{HD_intro}. Content addressable memories (CAMs) have become popular hardware solutions to perform similarity search due to their ability to perform parallel in-memory search \cite{HD_intro, fefet_mann, iMARS}. 

\par
A common method to perform similarity search with CAMs, \cite{narla2024cross} is to use the discharge time of the matchline (ML) as a measure of similarity. In these designs, each cell is connected to the ML through a discharge transistor that is turned on if there is a mismatch between the stored and search bits. As a result, the voltage on the precharged ML drops faster as the number of mismatching bits in a row increases. This approach is well-suited for large dataset searches, as a simple inverter can serve as a sense amplifier for each row, eliminating the need for large, power-hungry sense amplifiers. However, there are major issues that limit the resolution and accuracy of the search operation that our proposal aims to address 1) The time to turn on the discharge transistors in a row can vary significantly depending on the distance of the row to the search-line drivers because of the RC delay and IR drop associated with interconnects, 2) The discharge current generated by cells can also vary across the array because of the device and interconnect variabilities, and 3) As the Hamming distance (HDist) increases, the discharge time becomes too small, making it difficult to distinguish between delays corresponding to various large HDists. To improve the distinguishability between various HDists, it is desired to lower the discharge current of each cell. However, this would require near- or sub-threshold operation which makes the discharge current too sensitive to threshold voltage variability. As we move towards more advanced nodes and as the array size increases, these challenges grow and they can limit the array size and/or the resolution and accuracy of the search operations. 
\par
To mitigate these issues, we propose a design modification to the CAMs in \cite{narla2024cross, narla2022design, yin2018ultra, pagiamtzis2006content} by adding a fixed, linear, and relatively large and low-variability resistor to the discharge path of each cell as shown in Fig \ref{fig_pre_sch_bench}. By choosing an adequately large resistance, discharge time gets determined predominantly by this resistance value rather than the on-current of the transistors that might vary significantly across the array due to device variabilities and wire parasitics. Additionally, we propose a prolonged precharge scheme enabled by the reference resistance in our modified CAM that further enhances the robustness and variation-tolerance of our design by mitigating the effect of RC parasitics on the discharge delay.

\par
Fabricating resistors in integrated circuits with large and precise values especially at scaled technology nodes is challenging due to process variability, line edge roughness, and material inconsistencies. Quantum effects such as electron tunneling, increased temperature sensitivity, and variations in contact resistance further contribute to unpredictable resistance values. Here, we present successful fabrication of resistive devices that have relatively constant resistance values for the voltages of interest, and have the potential to be realized in the compact geometries needed for advanced CMOS nodes.

\par
The rest of this letter is organized as follows. The fabrication process and the experimental results for reference resistances are presented in Section II. Section III reviews the CAM cell design and the modeling framework for CAM arrays. The results in terms of resolution, sensitivity to variability, energy/delay are presented in Section IV, and Section V summarizes the letter.

\section{Resistive Elements}

\begin{figure}[t]
	\centering
	\includegraphics[width=0.95\linewidth]{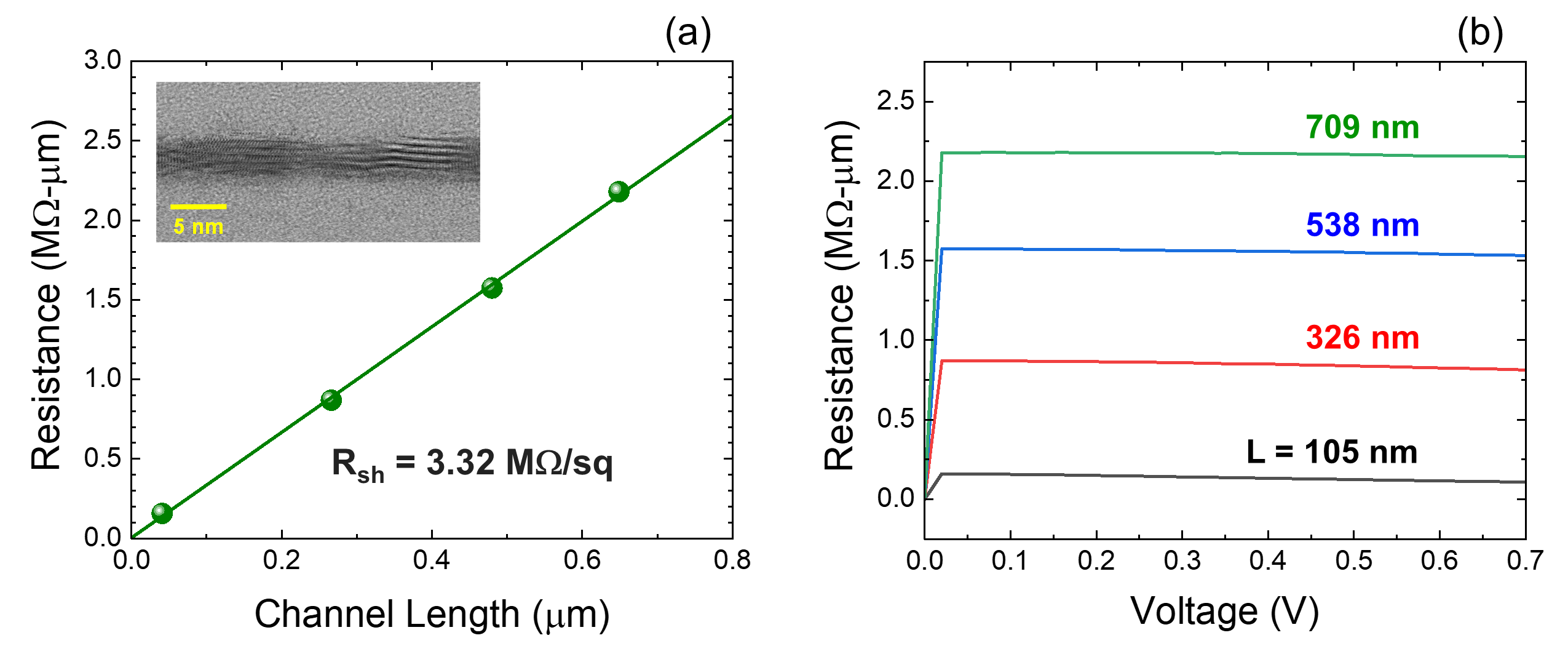}
	\caption{(a) TLM data and sheet resistance extraction of ALD TMD material. Inset: cross-section transmission electron microscopy image of ALD Nb-doped WS\textsubscript{2} (b) Resistance vs. voltage for resistors with varying contact spacings.}
	\label{fig_sotr_exp_results}
\end{figure}

\par
The resistive elements analyzed as part of this work were realized using moderately-doped thin-film transition metal dichalcogenides (TMDs) deposited by atomic layer deposition (ALD). ALD TMDs are ideal for compact thin-film resistors due to their ability to have controlled doping, and precise thickness control. The polycrystalline nature of the films also leads to low mobility at low doping levels, ensuring that high resistance values can be achieved within a small area.
\par
In this work, Nb-doped WS\textsubscript{2} films, similar to those described in \cite{schulpen2024nb}, were utilized. These films had a thickness of 3-nm and Nb (W) concentration of 8\% (92\%) and were deposited by ALD at 350 °C. To create the resistors, the films were deposited on Si/SiO\textsubscript{2} substrates and patterned into a rectangular transfer length method (TLM) geometry with Pd/Au used as the contacts with spacing ranging from 0.1 to 0.7 $\mu$m. The TLM results (Fig. 1(a)) show that films had sheet resistivity of 3.3 M$\Omega$ /square \cite{li2025ultra}. The high resistivity is a result of the polycrystalline material (Fig. 1 (a) inset) which has 0.08 cm\textsuperscript{2}/Vs mobility and 8.4 × $10^{19}$ cm-3 hole concentration. It is notable that the sheet resistivity is very high relative to the contact resistance, minimizing contact-induced non-uniformities. Fig. 1(b) shows the resistance vs. voltage plots, respectively for the devices with different contact spacing. The plot shows excellent resistance uniformity within a range of voltages (0.05-0.7V) needed for CAM operation.  While these test structures had width of 10 $\mu$m, assuming the resistor can be scaled to a width of 0.2 $\mu$m, then a device with 0.2 $\mu$m finger spacing would produce a resistance of 3 M$\Omega$, and only consume an area of roughly 0.08 $\mu$$m^2$, assuming contacts with 0.1 $\mu$m width. The use of this approach to create the selector resistor is attractive for CAM integration given the BEOL-compatible nature of the ALD process.  
\section{Cell Design and Modeling Framework}
\begin{figure}[t]
	\centering
	\includegraphics[width=0.95\linewidth]{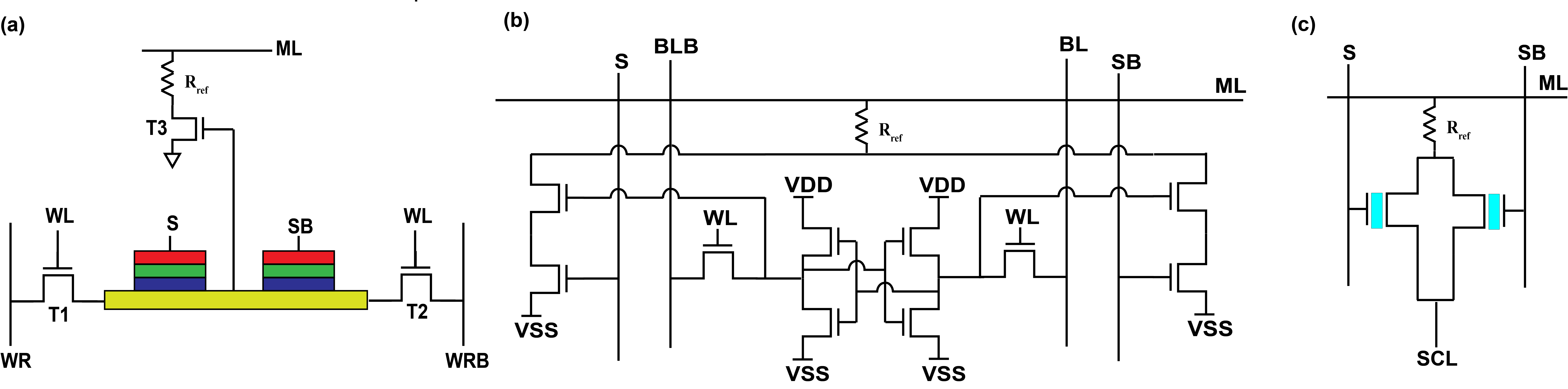}
	\caption{(a) SOT-R (b) SRAM-R and (c) FeFET-R CAM cell schematics}
	\label{fig_pre_sch_bench}
\end{figure}

Fig \ref{fig_pre_sch_bench} illustrates the modified designs for a spin-orbit-torque (SOT), ferroelectric field effect transistor (FeFET) and SRAM-based CAM referred to as SOT-R, FeFET-R and SRAM-R CAMs. In all three cases, complementary data is stored in a CAM cell, matchlines are precharged to Vdd before evaluation and during search, the searchlines, S and SB, are driven to Vs/0 depending on the search data.

\par
To perform accurate search simulations we have created physical layouts using ASAP7 \cite{clark2016asap7} and have extracted RC parasitics using Synopsis StarRC in the form of SPICE netlists. For SOT-CAMs we use MTJs with 45nm diameter and 2nm oxide thickness. The tunnel magnetoresistance (TMR), resistance-area product and half bias values are obtained from the experimental results in \cite{yuasa2004giant}. The TMR degradation due to bias voltage is also incorporated as shown in \cite{narla2022design}. For the FeFET-CAM, we use an FeFET with a memory window of 0.46V \cite{choe2021variability}. More details about the modeling framework, operation and performance of these CAMs are provided in \cite{narla2024cross}.

\begin{figure}[t]
	\centering
	\includegraphics[width=0.95\linewidth]{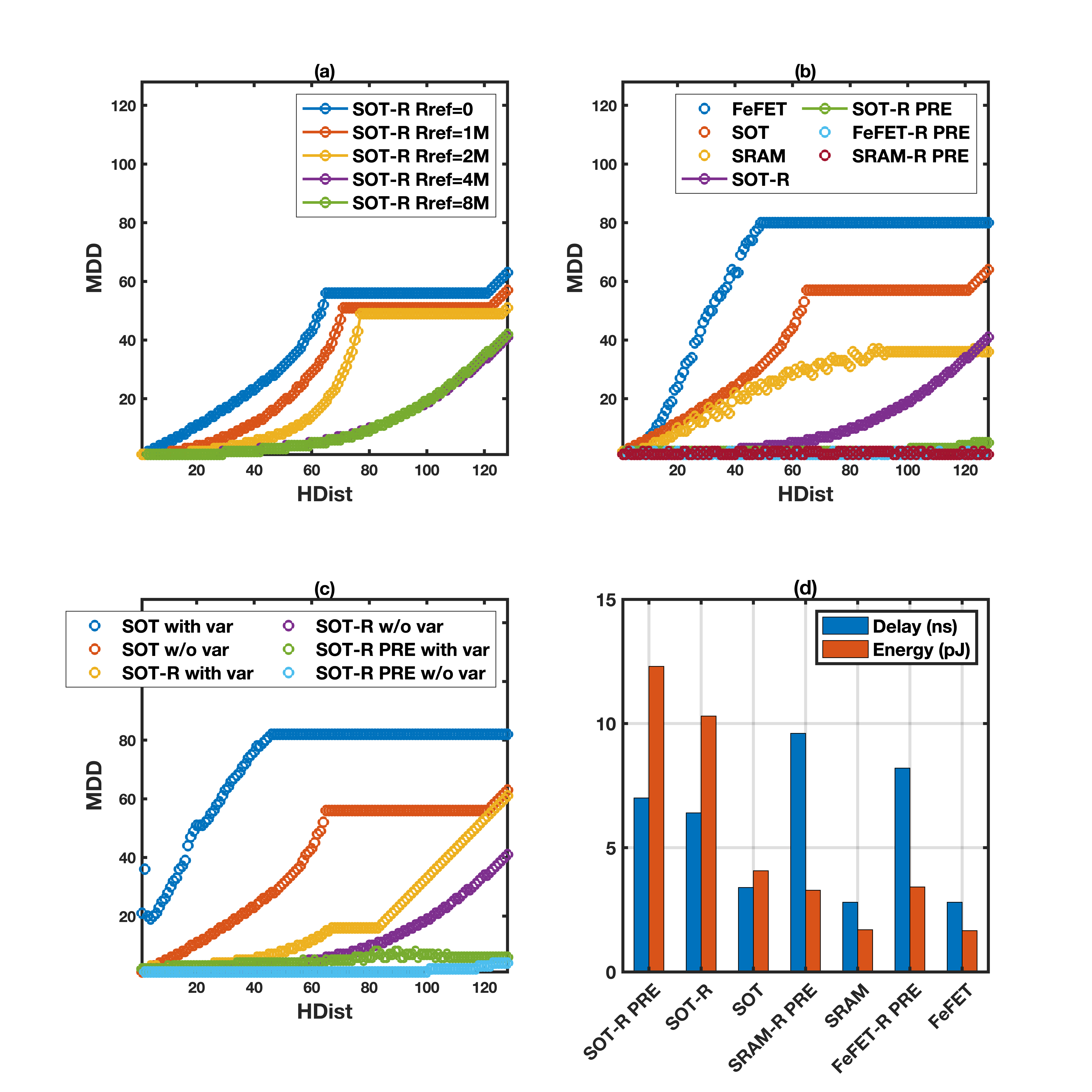}
	\caption{(MDD as a function of HDist for a 128x128 (a) SOT-R CAM array with various reference resistances, (b) SOT, FeFET and SRAM-based CAM arrays with and without modifications, (c) SOT, SOT-R and SOT-R PRE CAMs with and without variation (var), and (d) Search Energy and Delay results for HDist=20. }
	\label{fig_sotr_results}
\end{figure}

\section{Results}

\subsection{Resolution with Nominal Devices}

\par
We measure search resolution using the minimum detectable distance (MDD), as defined in \cite{narla2024cross}. The main factor limiting MDD is that a row close to the search lines’ drivers with a smaller HDist may discharge its ML faster than a row with a larger HDist that is farther away due to the effect of RC parasitics. Fig \ref{fig_sotr_results} (a) shows the impact of reference resistance value on the resolution of an SOT-R CAM array. As the resistance value increases, resolution improves for two main reasons. First, the overall discharge time increases, making it easier to distinguish between the delays of different HDists. Second, the RC product associated with the resistance of the reference resistor (R\textsubscript{ref}) and the capacitance of the ML dominates the overall delay. However, there is a diminishing return beyond 4M $\Omega$. Despite the use of reference resistors, there is variability in the overall discharge delays due to the variation introduced by the RC delays to turn on the discharge transistors and the IR drop across the search lines. Transistor threshold voltage and MTJ resistance variabilities further aggravate ML discharge delay variability. 

\par
To make our design insensitive to the above-mentioned variabilities, we propose to hold the ML at Vdd for a period of time after applying the search voltages to the searchlines. During this time, all the nodes across the array would reach their steady state values and the discharge transistors of all mismatching cells will be fully turned on regardless of their position in the array or their threshold voltage, and their discharge current becomes fully dependent on the resistance of R\textsubscript{ref}. At that point in time, the drivers holding MLs to Vdd are turned to the high-impedance mode and the search operation begins. Fig. \ref{fig_sotr_results} (b) show MDD results when the prolonged precharge technique is used for the SOT (SOT-R PRE), FeFET (FeFET-R PRE) and SRAM (SRAM-R PRE) CAM designs in the presence of R\textsubscript{ref}. For the FeFET and SRAM-CAMs in the absence of device variabilities, delay variabilities across the array exist solely due to RC delay in the circuit. While for the SOT-CAMs, along with RC delay, IR drop also contributes to delay variability due to the large current flowing through the searchlines. The prolonged precharge scheme eliminates RC delay but it remains slightly affected by IR drops. Hence, while we achieve state-of-the-art resolutions ($\leq$ 5-bits) for all three CAM designs, the FeFET and SRAM-CAMs benefit more from the prolonged precharge scheme than the SOT-CAMs. 
\par
This prolonged precharge strategy cannot be applied to the original CAM designs due to the prohibitive energy consumption. However, the large reference resistance limits the current passing through the MLs during this prolonged precharge period and the energy penalty corresponding to this approach is acceptable as will be reported in Section IV-C.

\subsection{Variability Study}
\par
The addition of the reference resistance and the prolong pre-charge are also effective tools in mitigating the effects of device variabilities such as changes in threshold voltage and MTJ diameter. We ran 100 Monte Carlo simulations for each HDist for the SOT-based CAMs. We used 27\%, 15\% and 42mV 3-sigma values for reference resistance (from our studies in Section II), MTJ resistance \cite{everspin} and threshold voltage \cite{giles2015high} variation, respectively. We used the data from the Monte Carlo simulations and fit it to a Gaussian distribution. We used the 3-sigma corners for each Hamming distance's delay distribution to calculate MDD. 
Fig \ref{fig_sotr_results} (c) shows that the SOT-R and SOT-R PRE-CAMs are more variation-tolerant due to the reduced dependence of discharge delay on the gate voltage of the discharge transistor which is susceptible to various sources of device variabilities.

\subsection{Energy and Delay Results}

\par
Adding R\textsubscript{ref} to the discharge path slows down the ML discharge and hence increases the search delay as compared to the previous CAM designs as seen in Fig \ref{fig_sotr_results} (d). Due to the larger discharge delay, search energy for the modified CAMs is larger. The prolonged precharge scheme requires additional energy to hold the ML at Vdd after driving the searchlines. This additional energy is not significant due to the large resistance attached to the ML. The delay in CAM-based similarity searches remains minimal, even for extremely large datasets. In contrast, most traditional algorithms \cite{luo2023survey, wang2021comprehensive} struggle to scale efficiently with increasing dataset sizes.

\section{Conclusion}
In this paper we have presented an updated CAM design with a fixed resistive device in the discharge path at the 7nm technology node. We showcased a novel BEOL compatible ALD of thin films and used the sheet resistance to realize a fixed resistor. To further benefit from the addition of the fixed resistor, we proposed a prolonged precharge scheme. Using this scheme, we observed near-ideal resolution ($\leq$ 5-bits) and achieved greater variation-tolerance to device variabilities at the cost of a 3x (2x) increase in energy (delay).

\bibliographystyle{unsrt}
\bibliography{sample-base}

\end{document}